\documentclass[aps,prl,twocolumn,footinbib]{revtex4-2}
\usepackage{amsmath,amsfonts,amssymb}
\usepackage{graphicx,float,calc}
\usepackage{color,bm}
\usepackage[colorlinks,urlcolor=blue,citecolor=blue,linkcolor=blue]{hyperref}
\usepackage{mathrsfs}
\usepackage[sort&compress]{natbib}
\usepackage{bbold}
\usepackage{listings}

\begin{document}

\title{Getting topological invariants from snapshots: a protocol for
defining and calculating topological invariants of systems with discrete
parameter space}
\author{Youjiang Xu, Walter Hofstetter}

\affiliation{Goethe-Universit\"at, Institut f\"ur Theoretische Physik, 60438
Frankfurt am Main, Germany}

\begin{abstract}
Topological invariants, including the Chern numbers, can topologically
classify parameterized Hamiltonians. We find that topological invariants can
be properly defined and calculated even if the parameter space is discrete,
which is done by geodesic interpolation in the classifying space. We
specifically present the interpolation protocol for the Chern numbers, which
can be directly generalized to other topological invariants. The protocol
generates a highly efficient algorithm for numerical calculation of the
second and higher Chern numbers, by which arbitrary precision can be
achieved given the values of the parameterized Hamiltonians on a coarse grid with a fixed resolution in the parameter space. 
Our findings also open up opportunities to study topology in finite-size systems where the parameter space can be naturally discrete.
\end{abstract}

\maketitle

Topological invariants can indicate whether it is possible to continuously
deform two manifolds into each other \cite{nakahara2018geometry}. They
appear in various systems, classifying parameterized Hamiltonians or their
eigenstates \cite{RevModPhys.82.3045, RevModPhys.83.1057,
RevModPhys.88.021004, RevModPhys.80.1083, RevModPhys.91.015005,
RevModPhys.89.041004, RevModPhys.91.015006, RevModPhys.93.015005,
ozawa2019topological}. For example, it is well known that in the Integer
Quantum Hall Effect, the Hall conductivity can only take integer values
because it is proportional to the 1st Chern number, a topological invariant
characterizing how the two-dimensional Brillouin zone is mapped to
single-particle Hamiltonians in momentum space \cite{PhysRevLett.49.405,
KOHMOTO1985343, PhysRevLett.71.3697}. The integer-valued 2nd and higher
Chern numbers can also show up as electromagnetic response \cite%
{PhysRevB.78.195424, PhysRevB.98.094434} or in quasicrystals \cite%
{PhysRevLett.111.226401, PhysRevResearch.4.013028, PhysRevB.105.115410}.
However, it is challenging to numerically calculate these quantities from
the Hamiltonians.\ This difficulty is illustrated by the following example.

Consider the Chern numbers of the family of single-particle Hamiltonians $%
H\left( \mathbf{k}\right) $ parameterized by a vector $\mathbf{k}\in K$,
where $K\equiv \left[ 0,1\right] ^{D}$ is the $D$-dimensional parameter
space. $H\left( \mathbf{k}\right) $ is an $s\times s$ hermitian matrix that
is periodic in each component $\left( k^{1},k^{2},\dots ,k^{D}\right) $ of $%
\mathbf{k}$ with period $1$, and it can also be regarded as a mapping from $K
$ to $s\times s$ hermitian matrices. Suppose $H\left( \mathbf{k}\right) $
describes a fermionic topological insulator gapped at zero energy for all $%
\mathbf{k}$, i.e., the number of negative eigenvalues of $H\left( \mathbf{k}%
\right) $, denoted as $r$, is independent of $\mathbf{k}$. We represent the $%
r$ lower-energy eigenstates of $H\left( \mathbf{k}\right) $ by an $s\times r$
matrix, denoted as $V\left( \mathbf{k}\right) $, where each column of $%
V\left( \mathbf{k}\right) $ is one of these normalized eigenstates occupied
by a fermion. $V\left( \mathbf{k}\right) $ satisfies $V^{\dag }\left(
\mathbf{k}\right) V\left( \mathbf{k}\right) =1$, and the collection of $%
s\times r$ matrices satisfying this relation forms the (complex)\ Stiefel
manifold $\mathrm{St}\left( s,r\right) \equiv \frac{U\left( s\right) }{%
U\left( s-r\right) }$, where $U\left( \cdot \right) $ represents the unitary
group. Then, the matrix forms of the Berry connection and the Berry
curvature are given by $\mathcal{A}_{\mu }\equiv V^{\dag }\partial _{\mu }V$
and $\mathcal{F}_{\mu \nu }=\partial _{\mu }\mathcal{A}_{\nu }-\partial
_{\nu }\mathcal{A}_{\mu }+\left[ \mathcal{A}_{\mu },\mathcal{A}_{\nu }\right]
$, respectively, where $\partial _{\mu }\equiv \frac{\partial }{\partial
k^{\mu }}$. The $n$th Chern character $\mathrm{ch}_{n}\equiv \frac{\epsilon
^{\mu _{1}\lambda _{1}\mu _{2}\lambda _{2}\cdots \mu _{n}\lambda _{n}}}{n!}%
\left( \frac{i}{4\pi }\right) ^{n}\mathrm{tr}\left[ \mathcal{F}_{\mu _{1}\nu
_{1}}\mathcal{F}_{\mu _{2}\nu _{2}}\cdots \mathcal{F}_{\mu _{n}\nu _{n}}%
\right] $, where $\epsilon ^{\mu _{1}\lambda _{1}\mu _{2}\lambda _{2}\cdots
\mu _{n}\lambda _{n}}$ is the Levi-Civita symbol, is integrated to give the $%
n$th Chern number $c_{n}$, i.e., ${c_{n}\equiv \int \mathrm{ch}_{n}\,\mathrm{%
d}^{2n}k}$, where the domain of the integral is a certain $2n$-dimensional
subspace of $K$, usually $K$ itself. Both $c_{n}$ and $\mathrm{ch}_{n}$ are
functionals of $V$, so they can be denoted as $c_{n}\left[ V\right] $ and $%
\mathrm{ch}_{n}\left[ V\right] $, respectively. Numerically, we may
straightforwardly follow these defining expressions to calculate the Chern
numbers: First, we discretize the parameter space $K$ by the grid $%
K_{0}\equiv \left\{ 0,N_{0}^{-1},2N_{0}^{-1},\dots ,1\right\} ^{D}$, which
divides $K$ into $N_{0}$ (integer)$\ $segments in each dimension. Then, for
any point on the grid, $\boldsymbol{\kappa }\in K_{0}$, we get the
Hamiltonian $H\left( \boldsymbol{\kappa }\right) $, diagonalize it to obtain
$V\left( \boldsymbol{\kappa }\right) $, from which we approximate $\mathcal{F%
}_{\mu \nu }\left( \boldsymbol{\kappa }\right) $ by finite difference
methods, and finally calculate the sum $\sum_{\boldsymbol{\kappa }}\mathrm{ch%
}_{n}\left( \boldsymbol{\kappa }\right) $ to get $c_{n}$. The precision of
such algorithms depends on the grid density $N_{0}$. For example, the
authors \cite{MocholGrzelak2018} follow these steps to calculate $c_{2}$,
and their algorithm is claimed to be efficient because they found a good way
to approximate $\mathcal{F}_{\mu \nu }\left( \boldsymbol{\kappa }\right) $
using $V\left( \boldsymbol{\kappa }\right) $. However, the algorithm in \cite%
{MocholGrzelak2018} still requires a large $N_{0}$, e.g. $60$, to reach
convergence, i.e. the algorithm requires $60^{4}$ Hamiltonians $H\left(
\boldsymbol{\kappa }\right) $ to be diagonalized, which can be too costy,
especially when $H\left( \boldsymbol{\kappa }\right) $ are effective
topological Hamiltonians \cite{PhysRevX.2.031008, Wang2013} that
characterizes topology of interacting systems, and obtaining a single $%
H\left( \boldsymbol{\kappa }\right) $ requires a significant amount of
computing resources. Another problem of such algorithms is that they cannot
efficiently pinpoint the topological phase boundaries where the topological
invariant $c_{n}$ jumps. This is because near the boundaries, $\mathrm{ch}%
_{n}\left( \mathbf{k}\right) $ peaks in small regions of $K$, and it is very
inefficient to capture the peak profile by increasing $N_{0}$.

These problems are avoided in the widely used algorithm calculating $c_{1}$
\cite{Fukui2005}. The authors of \cite{Fukui2005}~show that the exact value
of $c_{1}$ can be obtained with a fixed small $N_{0}$. In this paper, we
show that this is true for any Chern number $c_{n}$, which is achieved by a
proper \textit{ab initio} definition of the Chern numbers in a discrete
parameter space following a protocol of interpolation. We will use the
algorithm arising from this new protocol to calculate $c_{2}$ for the
lattice Dirac model and show its significant advantages in comparison to the
previous algorithm in \cite{MocholGrzelak2018}. Also, our protocol can
reproduce and provide an explanation for the existing algorithm for $c_{1}$
in \cite{Fukui2005}, and can be directly generalized to other topological
invariants besides Chern numbers.

The basic idea behind the protocol can be illustrated as follows: Consider
an object periodically moving around a circle with its trajectory described
by the azimuth $\theta \left( t\right) $, a continuous function of time $t$
satisfying the periodic boundary condition $\theta \left( t+1\right) =2\pi
w+\theta \left( t\right) $, where the winding number $w$, which counts the
number of circles the object completes in a period, is the topological
invariant that classifies the motion or the mapping $\theta \left( t\right) $%
. We can measure $\theta \left( t\right) $ to get $w$. In practice, it is
impossible to measure $\theta \left( t\right) $ at every moment $t$,
instead, we measure $\theta \left( t_{i}\right) $ at a finite number of
times $t_{1}$, $t_{2}$, $\dots $, $t_{N}$, supposing $t_{N}=t_{1}+1$, so the
information at hand is a discrete mapping $\theta \left( t_{i}\right) $.
Notice that we can determine $\theta \left( t_{i}\right) $ only up to modulo
of $2\pi $, because $\theta $ and $\theta +2\pi $ represent the same
physical position. If we assume that $\theta \left( t_{i}\right) $ is
measured so frequently that the object never moves more than half a circle
between adjacent measurements, we can eliminate the uncertainty of $\theta
\left( t_{i+1}\right) -\theta \left( t_{i}\right) $. For example, if $\theta
\left( t_{1}\right) =0$ and $\theta \left( t_{2}\right) =2\pi m+\delta $
with an unknown integer $m$ and a small $\delta $ satisfying $\left\vert
\delta \right\vert <\pi $, then $m$ has to be $0$. In this way, we obtain
the difference $\theta \left( t_{N}\right) -\theta \left( t_{1}\right) $ as
well as the winding number. For more complicated systems, topological
invariants can rarely be obtained from a simple subtraction like this, but
from certain defining integrals, e.g., the Chern numbers are integrals of
Chern characters. Such a defining integral for $w$ is $w=\int_{0}^{1}\frac{%
\mathrm{d}t}{2\pi }\frac{\mathrm{d}\theta \left( t\right) }{\mathrm{d}t}$.
The assumption we made implies that the same winding number can also be
obtained from $w=\int_{0}^{1}\frac{\mathrm{d}t}{2\pi }\frac{\mathrm{d}\tilde{%
\theta}\left( t\right) }{\mathrm{d}t}$, where $\tilde{\theta}\left( t\right)
\equiv \frac{\left( t_{i+1}-t\right) \theta \left( t_{i}\right) +\left(
t-t_{i}\right) \theta \left( t_{i+1}\right) }{t_{i+1}-t_{i}}$ with $t\in %
\left[ t_{i},t_{i+1}\right] $ and $\tilde{\theta}\left( t\right) $ is
defined for all intervals $\left[ t_{i},t_{i+1}\right] $ with any $i$. $%
\tilde{\theta}\left( t\right) $ is a continuous function in $\left[ 0,1%
\right] $, but its derivative $\frac{\mathrm{d}\tilde{\theta}\left( t\right)
}{\mathrm{d}t}$ is discontinuous at $t=t_{i}$, which does not make the
integral $\int_{0}^{1}\frac{\mathrm{d}t}{2\pi }\frac{\mathrm{d}\tilde{\theta}%
\left( t\right) }{\mathrm{d}t}$ ill-defined. Notice that in the interval $%
\left[ t_{i},t_{i+1}\right] $, $\tilde{\theta}\left( t\right) $ is one of
the geodesics that interpolates $\theta \left( t_{i}\right) $ and $\theta
\left( t_{i+1}\right) $, in the sense that the distance between $\theta
\left( t_{i}\right) $ and $\theta \left( t_{i+1}\right) $ given by $%
\int_{t_{i}}^{t_{i}+1}\left\vert d\tilde{\theta}\right\vert $ is its minimum
$\left\vert \theta \left( t_{i}\right) -\theta \left( t_{i}+1\right)
\right\vert $. Meanwhile, any function $\tilde{\theta}\left( \lambda \left(
t\right) \right) $, where $\lambda \left( t\right) $ satisfies $\lambda
\left( t_{i}\right) =t_{i}$, $\lambda \left( t_{i+1}\right) =t_{i+1}$ and $%
\frac{\mathrm{d}\lambda \left( t\right) }{\mathrm{d}t}>0$, represents the
same geodesic, and replacing $\tilde{\theta}\left( t\right) $ by $\tilde{%
\theta}\left( \lambda \left( t\right) \right) $ in $\int_{0}^{1}\frac{%
\mathrm{d}t}{2\pi }\frac{\mathrm{d}\tilde{\theta}\left( t\right) }{\mathrm{d}%
t}$ does not affect the winding number. This is because $\tilde{\theta}%
\left( t\right) $ and $\tilde{\theta}\left( \lambda \left( t\right) \right) $
are homotopic with each other (also with $\theta \left( t\right) $), i.e.,
they can be converted into each other by a continuous deformation. Because
the explicit form of $\theta \left( t\right) $ does not affect the winding
number calculated from $\tilde{\theta}\left( t\right) $, we can \textit{ab
initio} define the winding number of the discrete mapping $\theta \left(
t_{i}\right) $ by that of its geodesic interpolant $\tilde{\theta}\left(
t\right) $, without $\theta \left( t\right) $ coming into play. We will show
that the very idea of geodesic interpolation can also be used to define the
Chern numbers in discrete parameter spaces.

The goal is to find a proper definition of the Chern numbers knowing only
the discrete mapping $H\left( \boldsymbol{\kappa }\right) $ with $%
\boldsymbol{\kappa }\in K_{0}$. The definition only makes sense if the Chern
numbers of the discrete mapping $H\left( \boldsymbol{\kappa }\right) $
coincide with that of the continuous mapping $H\left( \mathbf{k}\right) $
given a large enough $N_{0}$. Provided the geodesic interpolants of $V\left(
\boldsymbol{\kappa }\right) $, denoted as $\tilde{V}\left( \mathbf{k}\right)
$, we define the Chern numbers of the mapping $H\left( \boldsymbol{\kappa }%
\right) $ as $c_{n}\left[ \tilde{V}\right] $, and we will show how to obtain
$\tilde{V}\left( \mathbf{k}\right) $ from $V\left( \boldsymbol{\kappa }%
\right) $.

Now, $V\left( \boldsymbol{\kappa }\right) $ plays a similar role as $\theta
\left( t_{i}\right) $ in the previous example. We know $\theta $ has the
uncertainty of modulo of $2\pi $, which however does not affect the winding
number. A similar situation occurs with any $V\in \mathrm{St}\left(
s,r\right) $. Physically, $V$ corresponds to a many-body state of $r$
fermions, the Slater determinant of the $r$ occupied states. The physical
many-body state is invariant under the gauge transformation represented by a
$r\times r$ unitary matrix $g$, i.e., $V$ and $Vg$ represent the same
physical state, which we denote by $V\simeq Vg$. As a result, Chern numbers
should not change under gauge transformations, i.e., $c_{n}\left[ V\right]
=c_{n}\left[ V^{\prime }\right] $ if $V\left( \mathbf{k}\right) \simeq
V^{\prime }\left( \mathbf{k}\right) $ for all $\mathbf{k}$. So what really
matters in determining the topology is the gauge invariant projection
operator $\rho \left( \mathbf{k}\right) \equiv V\left( \mathbf{k}\right)
V\left( \mathbf{k}\right) ^{\dag }$ that belongs to the (complex) Grassmann
manifold, ${\mathrm{Gr}\left( s,r\right) \equiv \frac{U\left( s\right) }{%
U\left( r\right) U\left( s-r\right) }}$, and an element in $\mathrm{Gr}%
\left( s,r\right) $ can be represented by different elements in $\mathrm{St}%
\left( s,r\right) $ that are related by gauge transformations. If we regard
a single-particle state as a vector in the space $%
\mathbb{C}
^{s}$, then $\rho \left( \mathbf{k}\right) $ projects this vector to the
hyperplane spanned by column vectors of $V\left( \mathbf{k}\right) $, so $%
\rho $ as well as its corresponding many-particle state can be geometrically
interpreted as this hyperplane, and $\mathrm{Gr}\left( s,r\right) $ is the
collection of such $r$-dimensional hyperplanes that contain the origin of $%
\mathbb{C}
^{s}$. This geometric interpretation helps to find the geodesics in $\mathrm{%
Gr}\left( s,r\right) $.

With the Riemannian metric, the distance between these hyperplanes is
measured by the angles between them \cite{wong1967differential,
berceanu1997geometry}. Consider two hyperplanes represented by $%
V_{0},V_{1}\in \mathrm{St}\left( s,r\right) $. The $r$ angles between $%
V_{0},V_{1}$ are defined by the singular-value decomposition
\begin{equation}
L\cos \Theta R^{\dag }\equiv V_{1}^{\dag }V_{0}
\end{equation}
where $L,R\in U\left( r\right) $, $%
\cos \Theta $ is a non-negative diagonal matrix, and the $r$ angles are the
diagonal elements of $\Theta $. Geometrically, the column vectors of $V_{0}R$
and $V_{1}L$ span the two hyperplanes, respectively, and the angle between
the $i$th column vector of $V_{0}R$, denoted as $R_{i}$, and the $i$th
column vector of $V_{1}L$, denoted as $L_{i}$, is the $i$th diagonal element
of $\Theta $, denoted as $\Theta _{i}$. If we rotate $R_{i}$ towards $L_{i}$
by the amount of $t\Theta _{i}$ for $t\in \left[ 0,1\right] $, and we
collect these rotated vectors for all $i$'s to span a new hyperplane as a
function of $t$, then the resulting hyperplane is the geodesic connecting
the two hyperplanes represented by $V_{0}$ and $V_{1}$, which can be
represented by
\begin{eqnarray}
\tilde{V}\left( t\right)  &=&V_{0}R\csc \Theta \sin \left[ \left( 1-t\right)
\Theta \right] R^{\dag }  \notag \\
&+&V_{1}L\csc \Theta \sin \left[ t\Theta \right] R^{\dag }  \label{1DInt}
\end{eqnarray}%
Notice that $\tilde{V}\left( t\right) $ is unique if and only if $\cos
\Theta $ is positive definite. If $\cos \Theta $'s diagonal contains zeros,
then the two hyperplanes are too far from each other to define a unique
geodesic, so Eq.~(\ref{1DInt}) is no longer valid. A special property of $%
\tilde{V}\left( t\right) $ is that its corresponding Berry connection
vanishes, i.e.,
\begin{equation}
\tilde{V}^{\dag }\left( t\right) \partial _{t}\tilde{V}\left( t\right) =0
\label{A0}
\end{equation}%
However, one can verify that ${\tilde{V}\left( 0\right) =V_{0}}$ and ${%
\tilde{V}\left( 1\right) =V_{1}LR^{\dag }\simeq V_{1}}$, i.e., a gauge
transformation is applied at $t=1$. We can "repair" the gauge by adopting
another representation of the same geodesic in $\mathrm{Gr}\left( s,r\right)
$, e.g., $\bar{V}\left( t\right) \equiv \tilde{V}\left( t\right) \exp \left[
t\ln \left( RL^{\dag }\right) \right] $. One can verify that $\bar{V}\left(
0\right) =V_{0}$ and $\bar{V}\left( 1\right) =V_{1}$, however, the Berry
connection of this new representation no longer vanishes, i.e., $\bar{V}%
^{\dag }\partial _{t}\bar{V}=\ln \left( RL^{\dag }\right) $. In the
Supplemental Materials \cite{[{The Supplemental Material includes: i. The
explicit procedure of a two-dimensional interpolation that reproduces the
known algorithm for the first Chern number; ii. The proof of the uniqueness
of the interpolant hyperplane; iii. The Python code that illustrates the
algorithm for the second Chern number and compares its performance against a
previously established method. }] SM}, we will show that $\bar{V}$ can be used to
reproduce the existing algorithm for calculating the 1st Chern number in a
discrete Brillouin zone \cite{Fukui2005}.

The geodesic in Eq.~(\ref{1DInt}) is a one-dimensional interpolant provided
two endpoints $V_{0}$ and $V_{1}$. From Eq.~(\ref{1DInt}), we can generate
the $D$-dimensional interpolant $\tilde{V}\left( \mathbf{k}\right) $ that
defines the Chern numbers. The grid $K_{0}$ divides the parameter space $K$
into $N_{0}^{D}$ $D$-dimensional hypercubes, and we use $C_{\boldsymbol{%
\kappa }}$ to denote the hypercube whose vertex closest to the origin is ${%
\boldsymbol{\kappa }\in K_{0}}$, i.e., if ${\mathbf{k}\in C_{\boldsymbol{%
\kappa }}}$ then its components satisfy ${\mathbf{\,}0\leq k^{i}-\kappa
^{i}\leq N_{0}^{-1}}$. To get ${\tilde{V}\left( \mathbf{k}\right) }$, we
recursively define $V_{\boldsymbol{\kappa }}^{\left( 0\right) }\left(
\mathbf{k}\right) $ as follows: Given an integer ${0<i\leq D}$, for $\mathbf{%
k}\in C_{\boldsymbol{\kappa }}$ that satisfies $k^{j}=$ $\kappa ^{j}$ or $%
\kappa ^{j}+N_{0}^{-1}$ for any $j<i$, $V_{\boldsymbol{\kappa }}^{\left(
i-1\right) }\left( \mathbf{k}\right) $ is defined as the right hand side of
Eq.~(\ref{1DInt}) with the following replacement: ${V_{0}\rightarrow V_{%
\boldsymbol{\kappa }}^{\left( i\right) }\left( P_{\boldsymbol{\kappa }%
,0}^{i}\left( \mathbf{k}\right) \right) }$, ${V_{1}\rightarrow V_{%
\boldsymbol{\kappa }}^{\left( i\right) }\left( P_{\boldsymbol{\kappa }%
,1}^{i}\left( \mathbf{k}\right) \right) }$, and ${t\rightarrow N_{0}\left(
k^{i}-\kappa ^{i}\right) }$ with the initial condition ${V_{\boldsymbol{%
\kappa }}^{\left( D\right) }\left( \mathbf{k}\right) \equiv V\left( \mathbf{k%
}\right) }$, where the projection operator $P_{\boldsymbol{\kappa },t}^{i}$
for $0\leq t\leq 1$ applied to $\mathbf{k}$ changes $k^{i}$ to $%
tN_{0}^{-1}+\kappa ^{i}$ while it keeps other components of $\mathbf{k}$
invariant. In these recursive procedures, we assume that at each time we use
Eq.~(\ref{1DInt}), $V_{0}$ and $V_{1}$ are close enough such that $\tilde{V}%
\left( t\right) $ is unique, which can be achieved with a large enough $N_{0}
$. In the Supplemental Material \cite{SM}, the recursive procedure is
explicitly shown for calculating the first Chern number in a two-dimensional
parameter space.

$V_{\boldsymbol{\kappa }}^{\left( 0\right) }\left( \mathbf{k}\right) $ is a
smooth function with well-defined derivatives in its domain $C_{\boldsymbol{%
\kappa }}$. We define $\tilde{V}\left( \mathbf{k}\right) $ piecewise as $V_{%
\boldsymbol{\kappa }}^{\left( 0\right) }\left( \mathbf{k}\right) $, i.e., $%
\tilde{V}\left( \mathbf{k}\right) \equiv V_{\boldsymbol{\kappa }}^{\left(
0\right) }\left( \mathbf{k}\right) $ if $\mathbf{k}\in C_{\boldsymbol{\kappa
}}$. This definition is ambiguous (multi-valued) on the boundaries of $C_{%
\boldsymbol{\kappa }}$, however, this ambiguity will not affect the Chern
numbers, because the hyperplane represented by $\tilde{V}\left( \mathbf{k}%
\right) $ is unique, i.e., given $\boldsymbol{\kappa }_{0}$ and $\boldsymbol{%
\kappa }_{1}$ such that $C_{\boldsymbol{\kappa }_{0}}\cap C_{\boldsymbol{%
\kappa }_{1}}$ is non-empty, $V_{\boldsymbol{\kappa }_{0}}^{\left( 0\right)
}\left( \mathbf{k}\right) \simeq V_{\boldsymbol{\kappa }_{1}}^{\left(
0\right) }\left( \mathbf{k}\right) $ for any $\mathbf{k}\in C_{\boldsymbol{%
\kappa }_{0}}\cap C_{\boldsymbol{\kappa }_{1}}$. The proof of this statement
is given in the Supplemental Material \cite{SM}. As a result, $\tilde{\rho}%
\left( \mathbf{k}\right) \equiv \tilde{V}\left( \mathbf{k}\right) \tilde{V}%
^{\dag }\left( \mathbf{k}\right) $ is a single-valued, continuous, but not
necessarily smooth function for $\mathbf{k}\in K$. The lack of smoothness
does not affect calculating the Chern numbers as long as we piecewise
evaluate the defining integral, i.e.,%
\begin{equation}
c_{n}\left[ \tilde{V}\right] \equiv \sum_{\boldsymbol{\kappa }}\int_{C_{%
\boldsymbol{\kappa }}}\mathrm{ch}_{n}\left[ V_{\boldsymbol{\kappa }}^{\left(
0\right) }\left( \mathbf{k}\right) \right] \,\mathrm{d}^{2n}k
\label{piecewise}
\end{equation}%
This statement can be verified by showing that the Chern number defined by
Eq.~(\ref{piecewise}) coincides with $c_{n}\left[ V\left( \mathbf{k}\right) %
\right] $ given a large enough $N_{0}$. By increasing $N_{0}$, the distance
between $V\left( \mathbf{k}_{1}\right) $ and $V\left( \mathbf{k}_{2}\right) $
decreases for any ${\mathbf{k}_{1},\mathbf{k}_{2}\in C_{\boldsymbol{\kappa }}}$%
. For a large $N_{0}$, such distances are so small that there exists a
unique geodesic, represented by ${v\left( t,\mathbf{k}\right)}$ with $t\in %
\left[ 0,1\right] $, connecting $V\left( \mathbf{k}\right) $ and $V_{
\boldsymbol{\kappa }}^{\left( 0\right) }\left( \mathbf{k}\right) $, i.e., ${
v\left( 0,\mathbf{k}\right) \simeq V\left( \mathbf{k}\right)}$ and ${v\left(
1,\mathbf{k}\right) \simeq V_{\boldsymbol{\kappa }}^{\left( 0\right) }\left(
\mathbf{k}\right)}$. Now, $v\left( t,\mathbf{k}\right) $ establishes a
continuous deformation (homotopy) between $V\left( \mathbf{k}\right) $ and $%
V_{\boldsymbol{\kappa }}^{\left( 0\right) }\left( \mathbf{k}\right) $, which
guarantees that Eq.~(\ref{piecewise}) equals to $c_{n}\left[ V\left( \mathbf{%
k}\right) \right] $.

In conclusion, given the Hamiltonians $H\left( \boldsymbol{\kappa }\right) $
where $\boldsymbol{\kappa }$ belong to a finite-size ($N_{0}^{D}$) grid in
the parameter space $K$, we can define and calculate to arbitrary precision
the Chern numbers, which coincide with the Chern numbers of the continuous
mapping $H\left( \mathbf{k}\right) $ with $\mathbf{k}\in K$. Our result has
two inferences: 1. It provides a highly efficient algorithm to numerically
calculate the Chern numbers; 2. For systems with a discrete parameter space,
e.g., a finite-size tight-binding model with periodical boundary conditions,
our method provides an \textit{ab initio} definition of Chern numbers, which
may open up a plethora of new opportunities in topological physics.
Topological invariants other than the Chern numbers might have a different
classifying space than the Grassmann manifold \cite{RevModPhys.88.035005}.
For example, if the single-particle Hamiltonian $h\left( \mathbf{k}\right) $
is a $2s\times 2s$ hermitian matrix with chiral symmetry, its corresponding
many-particle ground state is no longer described by a matrix $V\left(
\mathbf{k}\right) $ in the Stiefel manifold, but is described by a unitary
matrix $q\left( \mathbf{k}\right) \in U\left( s\right) $, and if $q\left(
\mathbf{k}\right) $ is regarded as mapping from the parameter space to $%
U\left( s\right) $, it can be topologically classified by a winding number
\cite{RevModPhys.88.035005}. To define and calculate the winding number of
the discrete mapping $q\left( \boldsymbol{\kappa }\right) $ where $%
\boldsymbol{\kappa }$ belongs to a grid, one should find the geodesic
interpolants $\tilde{q}\left( \mathbf{k}\right) $ of $q\left( \boldsymbol{%
\kappa }\right) $, which can be done following the same procedure as the one
for interpolating $V\left( \boldsymbol{\kappa }\right) $ with the only
modification that Eq.~(\ref{1DInt}) should be replaced by the equation of
geodesics in $U\left( s\right) $, i.e., $q_{0},q_{1}\in U\left( s\right) $
are connected by $\tilde{q}\left( t\right) =q_{0}\exp \left( t\ln \left(
q_{0}^{\dag }q_{1}\right) \right) $.

\begin{figure}[tbh]
\centering
\includegraphics[width=0.5\textwidth]{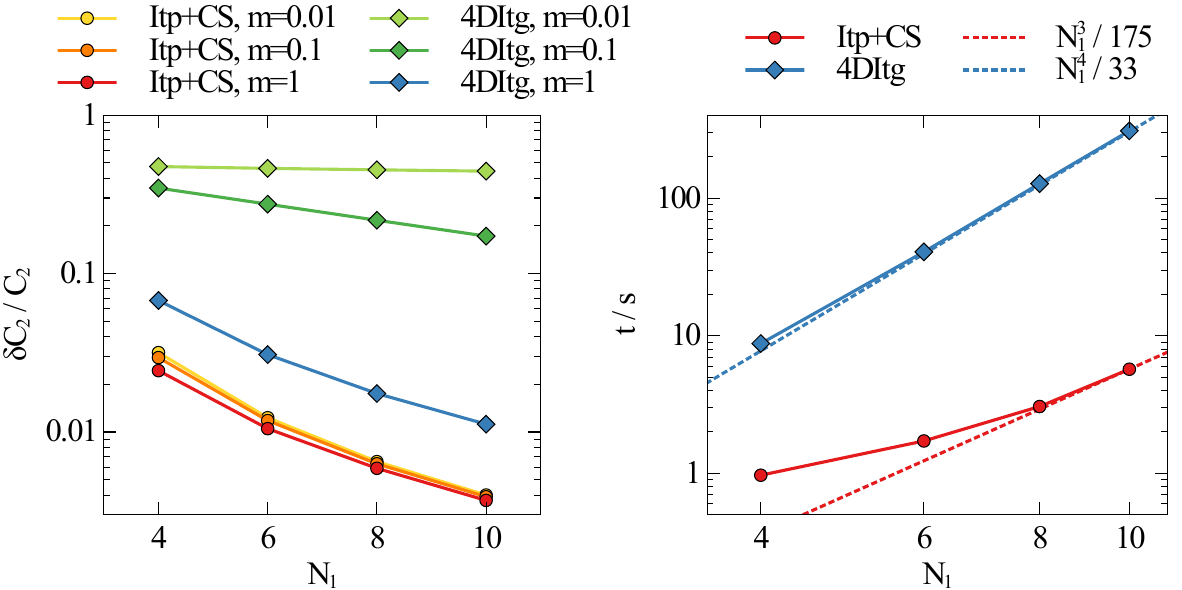}
\caption{ (Color online) We calculate the second Chern number of the Dirac
lattice model using both our new algorithm Itp+CS and the previously established one, 4DItg \cite
{MocholGrzelak2018}, with
varying grid density and compare (a) the relative error and (b) the time
cost. }
\label{Fig1}
\end{figure}

Finally, we provide additional information on the numerical implementation
as well as the benchmark of the algorithm for calculating the Chern numbers.
Instead of direct evaluation of Eq.~(\ref{piecewise}), the algorithm can be
further optimized. We assume $D=2n$ for simplicity. Direct numerical
evaluation of Eq.~(\ref{piecewise}) requires the values $V_{\boldsymbol{%
\kappa }}^{\left( 0\right) }\left( \mathbf{k}\right) $ for $\mathbf{k}$ on a
grid in $C_{\boldsymbol{\kappa }}$ consisting of $O\left( N_{1}^{D}\right) $
points, where $N_{1}$ is the number of different values of $t$ for which $%
\tilde{V}\left( t\right) $ in Eq.~(\ref{1DInt}) is evaluated. The larger $%
N_{1}$ is, the better the precision will be. Using the generalized Stokes'
theorem, the $D$-dimensional ``bulk'' integral $\int_{C_{\boldsymbol{\kappa }}}%
\mathrm{ch}_{n}\left[ V_{\boldsymbol{\kappa }}^{\left( 0\right) }\left(
\mathbf{k}\right) \right] \,\mathrm{d}^{D}k$ in Eq.~(\ref{piecewise}) can be
reduced to a $\left( D-1\right) $-dimensional ``surface'' integral, whose
integrand is the Chern-Simons form \cite{nakahara2018geometry,
RevModPhys.88.035005}, and now only $O\left( N_{1}^{D-1}\right) $ times of
evaluation of $V_{\boldsymbol{\kappa }}^{\left( 0\right) }\left( \mathbf{k}%
\right) $ is required. This is possible because $V_{\boldsymbol{\kappa }%
}^{\left( 0\right) }\left( \mathbf{k}\right) $ is a smooth function in $C_{%
\boldsymbol{\kappa }}$, even though $V\left( \mathbf{k}\right) $ is rarely
smooth in numerics. In particular, to calculate the second Chern number $%
c_{2}$ in our benchmark, we rewrite Eq.~(\ref{piecewise}) as
\begin{eqnarray}
c_{2}\left[ \tilde{V}\right]  &\equiv &-\frac{1}{8\pi ^{2}}\sum_{\boldsymbol{%
\kappa }}\int_{\partial C_{\boldsymbol{\kappa }}}\epsilon ^{\mu \nu \lambda
\delta }  \notag \\
&\mathrm{tr}&\left( \mathcal{A}_{\mu }^{\left( \boldsymbol{\kappa }\right)
}\partial _{\nu }\mathcal{A}_{\lambda }^{\left( \boldsymbol{\kappa }\right)
}+\frac{2}{3}\mathcal{A}_{\mu }^{\left( \boldsymbol{\kappa }\right) }%
\mathcal{A}_{\nu }^{\left( \boldsymbol{\kappa }\right) }\mathcal{A}_{\lambda
}^{\left( \boldsymbol{\kappa }\right) }\right) \mathrm{d}^{3}s_{\delta }
\label{c2}
\end{eqnarray}%
where $\mathcal{A}_{\mu }^{\left( \boldsymbol{\kappa }\right) }\equiv \left(
V_{\boldsymbol{\kappa }}^{\left( 0\right) }\right) ^{\dag }\partial _{\mu
}V_{\boldsymbol{\kappa }}^{\left( 0\right) }$ and $\mathrm{d}^{3}s_{\delta }$
is the 3-dimensional "surface" element in the direction $\delta $. As a
result of Eq.~(\ref{A0}), $\mathcal{A}_{1}^{\left( \boldsymbol{\kappa }%
\right) }=0$, which can provide further simplification of Eq.~(\ref{c2}). In
the Supplemental Material \cite{SM}, details of optimization are presented
in the Python code for the benchmark of the optimized algorithm, named
Itp+CS (Interpolation plus Chern-Simons Forms), and also a code for the algorithm used in \cite
{MocholGrzelak2018}, named 4DItg (4-dimensional Integration). Both algorithms are applied to
calculate the 2nd Chern number of the lattice Dirac model with the
Hamiltonian $H_{\mathrm{Dirac}}\left( \mathbf{k}\right) =\mathbf{d}\left(
\mathbf{k}\right) \,\mathbf{\cdot \,\Gamma }$ where $\mathbf{d}\left(
\mathbf{k}\right) $ has 5 components $d^{0}=m+\sum_{i=1}^{4}\cos 2\pi k_{i}$
and $d^{i}=\sin 2\pi k_{i}$ for $i=1,2,3,4$, and $\mathbf{\Gamma }$ are
Dirac matrices represented by products of Pauli matrices
\begin{equation*}
\mathbf{\Gamma }=\left( \sigma _{x}\otimes \mathbb{1},\sigma _{y}\otimes
\mathbb{1},\sigma _{y}\otimes \sigma _{x},\sigma _{y}\otimes \sigma
_{y},\sigma _{z}\otimes \sigma _{z}\right)
\end{equation*}%
The parameter $m$ determines the topological phase and there is a
topological phase boundary at $m=0$. In the benchmark, Itp+CS diagonalizes $%
H_{\mathrm{Dirac}}\left( \mathbf{k}\right) $ on the lattice $K_{0}$ with a
fixed $N_{0}=4$, and then interpolates the output with a varying $N_{1}$ to
get $V_{\boldsymbol{\kappa }}^{\left( 0\right) }\left( \mathbf{k}\right) $
for integrating the Chern-Simons form. Meanwhile, 4DItg diagonalizes $H_{%
\mathrm{Dirac}}\left( \mathbf{k}\right) $ on the lattice $K_{0}$ but with a
varying $N_{0}$, and we denote this varying $N_{0}$ as $N_{2}$ for
distinguishing it from the fixed one used in Itp+CS. By setting $%
N_{2}=N_{0}N_{1}$, we plot in Fig.~\ref{Fig1}(a) the relative error of the
2nd Chern number calculated by the two algorithms for different values of $m$
that approach the topological phase boundary $m=0$, and in Fig.~\ref{Fig1}%
(b) the typical time cost of the two algorithms. The results show that
Itp+CS always outperforms 4DInt in both precision and speed. Especially,
when it is close to a phase transition, i.e., $m=0.1$ or $0.01$, 4DItg fails
to converge while Itp+CS provides almost the same level of precision as it
does when $m=1$.

\begin{acknowledgments}
This work was supported by the Deutsche Forschungsgemeinschaft (DFG, German
Research Foundation) under Project No. 277974659 via Research Unit FOR 2414. The authors gratefully
acknowledge the Gauss Centre for Supercomputing e.V. (www.gauss-centre.eu) for funding this project
by providing computing time through the John von Neumann Institute for Computing (NIC) on the GCS
Supercomputer JUWELS at J\"ulich Supercomputing Centre (JSC). Calculations for this research were also
performed on the Goethe-NHR high performance computing cluster. The cluster is managed by the Center
for Scientific Computing (CSC) of the Goethe University Frankfurt.
\end{acknowledgments}


\newpage
\widetext
\begin{center}
\textbf{\large Supplemental Material - Getting topological invariants from
snapshots: a protocol for defining and calculating topological invariants of
systems with discrete parameter space}
\end{center}
\setcounter{equation}{0}
\setcounter{figure}{0}
\setcounter{table}{0}

\makeatletter
\renewcommand{\theequation}{S\arabic{equation}}
\renewcommand{\thefigure}{S\arabic{figure}}

\section{The algorithm for the 1st Chern number}

We show that the algorithm in \cite{Fukui2005} for calculating $c_{1}$ can
be reproduced from the geodesic interpolant $\bar{V}\left( t\right) \equiv
\tilde{V}\left( t\right) \exp \left[ t\ln \left( RL^{\dag }\right) \right] $
where the quantities on the r.h.s. of this equation are defined by Eq.~(1-2) in the main text.
Consider a two-dimensional ($D=2$) parameter
space $K$. Knowing $V\left( \boldsymbol{\kappa }\right) $ for $\boldsymbol{%
\kappa }$ belonging to the lattice $K_{0}$ in $K$, the two-dimensional
interpolant $\bar{V}\left( \mathbf{k}\right) $ is obtained in the same way
as $\tilde{V}\left( \mathbf{k}\right) $, with the exception that at each
time Eq.~(2) is used, we apply the additional gauge transformation
$\exp \left[ t\ln \left( RL^{\dag }\right) \right] $ to change $\tilde{V}%
\left( t\right) $ into $\bar{V}\left( t\right) $.

Let us illustrate these procedures in detail: Denote any $\boldsymbol{\kappa
}\in K_{0}$ by two integers $i_{1},i_{2}$ such that $\boldsymbol{\kappa }%
=\left( i_{1}/N_{0},i_{2}/N_{0}\right) $. In the hypercube $C_{\boldsymbol{%
\kappa }}$, first we have $V_{\boldsymbol{\kappa }}^{\left( 2\right) }\left(
\mathbf{k}\right) =V\left( \mathbf{k}\right) $, where $\mathbf{k}$ can only
be one of the vertices of $C\left( \boldsymbol{\kappa }\right) $, i.e., $%
\left( i_{1},i_{2}\right) /N_{0}$, $\left( i_{1}+1,i_{2}\right) /N_{0}$, $%
\left( i_{1},i_{2}+1\right) /N_{0}$, or $\left( i_{1}+1,i_{2}+1\right)
/N_{0} $. Next, we interpolate $V_{\boldsymbol{\kappa }}^{\left( 2\right)
}\left( \mathbf{k}\right) $ along the 2nd dimension to get $V_{\boldsymbol{%
\kappa }}^{\left( 1\right) }\left( \mathbf{k}\right) $. $V_{\boldsymbol{%
\kappa }}^{\left( 1\right) }\left( \mathbf{k}\right) $ has two disconnected
domains $k^{1}=i_{1}/N_{0}$ or $\left( i_{1}+1\right) /N_{0}$ and its
expression is%
\begin{eqnarray}
V_{\boldsymbol{\kappa }}^{\left( 1\right) }\left( k^{1},k^{2}\right) = &&V_{%
\boldsymbol{\kappa }}^{\left( 2\right) }\left( k^{1},i_{2}/N_{0}\right)
R^{\left( 2\right) }\csc \Theta ^{\left( 2\right) }\sin \left[ \left(
1-\left( N_{0}k^{2}-i_{2}\right) \right) \Theta ^{\left( 2\right) }\right]
R^{\left( 2\right) \dag }  \notag \\
+ &&V_{\boldsymbol{\kappa }}^{\left( 2\right) }\left( k^{1},\left(
i_{2}+1\right) /N_{0}\right) L^{\left( 2\right) }\csc \Theta ^{\left(
2\right) }\sin \left[ \left( N_{0}k^{2}-i_{2}\right) \Theta ^{\left(
2\right) }\right] R^{\left( 2\right) \dag }  \notag \\
= &&\bar{V}_{\boldsymbol{\kappa }}^{\left( 1\right) }\left(
k^{1},k^{2}\right) \exp \left[ -\left( N_{0}k^{2}-i_{2}\right) \ln \left(
R^{\left( 2\right) }L^{\left( 2\right) \dag }\right) \right]
\end{eqnarray}%
with
\begin{equation}
L^{\left( 2\right) }\csc \Theta ^{\left( 2\right) }R^{\left( 2\right) \dag
}\equiv \left[ V_{\boldsymbol{\kappa }}^{\left( 2\right) }\left(
k^{1},\left( i_{2}+1\right) /N_{0}\right) \right] ^{\dag }V_{\boldsymbol{%
\kappa }}^{\left( 2\right) }\left( k^{1},i_{2}/N_{0}\right)
\end{equation}%
Notice that $L^{\left( 2\right) }$, $R^{\left( 2\right) }$ and $\csc \Theta
^{\left( 2\right) }$ are functions of $k^{1}$, and we have applied the gauge
transformation, in contrast to the main text. Then we interpolate $\bar{V}_{%
\boldsymbol{\kappa }}^{\left( 1\right) }\left( k^{1},k^{2}\right) $ (in the
main text $V_{\boldsymbol{\kappa }}^{\left( 1\right) }\left(
k^{1},k^{2}\right) $)\ to get $\bar{V}_{\boldsymbol{\kappa }}^{\left(
0\right) }\left( k^{1},k^{2}\right) $ (in the main text $V_{\boldsymbol{%
\kappa }}^{\left( 0\right) }\left( k^{1},k^{2}\right) $)\ , whose domain is $%
C_{\boldsymbol{\kappa }}$
\begin{eqnarray}
&&\bar{V}_{\boldsymbol{\kappa }}^{\left( 0\right) }\left( k^{1},k^{2}\right)
\exp \left[ -\left( N_{0}k^{1}-i_{1}\right) \ln \left( R^{\left( 1\right)
}L^{\left( 1\right) \dag }\right) \right]  \notag \\
= &&\bar{V}_{\boldsymbol{\kappa }}^{\left( 1\right) }\left(
i_{1}/N_{0},k^{2}\right) R^{\left( 1\right) }\csc \Theta ^{\left( 1\right)
}\sin \left[ \left( 1-\left( N_{0}k^{1}-i_{1}\right) \right) \Theta ^{\left(
1\right) }\right] R^{\left( 1\right) \dag }  \notag \\
+ &&\bar{V}_{\boldsymbol{\kappa }}^{\left( 1\right) }\left( \left(
i_{1}+1\right) /N_{0},k^{2}\right) L^{\left( 1\right) }\csc \Theta ^{\left(
1\right) }\sin \left[ \left( N_{0}k^{1}-i_{1}\right) \Theta ^{\left(
1\right) }\right] R^{\left( 1\right) \dag }
\end{eqnarray}%
with
\begin{equation}
L^{\left( 1\right) }\csc \Theta ^{\left( 1\right) }R^{\left( 1\right) \dag
}= \left[ \bar{V}_{\boldsymbol{\kappa }}^{\left( 1\right) }\left( \left(
i_{1}+1\right) /N_{0},k^{2}\right) \right] ^{\dag }\bar{V}_{\boldsymbol{%
\kappa }}^{\left( 1\right) }\left( i_{1}/N_{0},k^{2}\right)
\end{equation}%
Notice that $L^{\left( 1\right) }$, $R^{\left( 1\right) }$ and $\csc \Theta
^{\left( 1\right) }$ are functions of $k^{2}$. We define $\bar{V}\left(
\mathbf{k}\right) $ piecewise as $\bar{V}_{\boldsymbol{\kappa }}^{\left(
0\right) }\left( \mathbf{k}\right) $, i.e., $\bar{V}\left( \mathbf{k}\right)
\equiv \bar{V}_{\boldsymbol{\kappa }}^{\left( 0\right) }\left( \mathbf{k}%
\right) $ if $\mathbf{k}\in C_{\boldsymbol{\kappa }}$.$\ $This definition is
not ambiguous because $\bar{V}\left( t\right) $, unlike $\tilde{V}\left(
t\right) $, conserves the gauge convention, i.e., $\bar{V}\left( 0\right)
=V_{0}$ and $\bar{V}\left( 1\right) =V_{1}$, see the discussion in the main
text after Eq.~(2). Nevertheless, $\bar{V}\left( \mathbf{k}\right)
$ and $\tilde{V}\left( \mathbf{k}\right) $ provide the \emph{same} geodesic
interpolant because $\bar{V}\left( \mathbf{k}\right) \simeq \tilde{V}\left(
\mathbf{k}\right) $ for all $\mathbf{k}$.

The 1st Chern number is given by%
\begin{eqnarray}
c_{1}\left[ \bar{V}\right]  &\equiv &\sum_{\boldsymbol{\kappa }}\int_{C_{%
\boldsymbol{\kappa }}}\frac{i}{2\pi }\epsilon ^{\mu \nu }\mathrm{tr}\left(
\mathcal{F}_{\mu \nu }\right) \,\mathrm{d}^{2}k  \notag \\
&=&\sum_{\boldsymbol{\kappa }}\oint\limits_{\partial C_{\boldsymbol{\kappa }%
}}\frac{i}{2\pi }\mathrm{tr}\left( \mathcal{A}_{\mu }\right) \,\mathrm{d}%
k^{\mu }  \notag \\
&=&\sum_{\boldsymbol{\kappa }}\oint\limits_{\partial C_{\boldsymbol{\kappa }%
}}\frac{i}{2\pi }\mathrm{tr}\left( \bar{V}^{\dag }\partial _{\mu }\bar{V}%
\right) \,\mathrm{d}k^{\mu }
\end{eqnarray}%
where in the 2nd line we have used Stokes' theorem to cast the
two-dimensional integral over $C_{\boldsymbol{\kappa }}$ into a contour
integral along the boundary of $C_{\boldsymbol{\kappa }}$, denoted as $%
\partial C_{\boldsymbol{\kappa }}$. $\partial C_{\boldsymbol{\kappa }}$ has
four segments that contribute to the integral in the same manner. For
example, consider the segment linking $\left( i_{1},i_{2}\right) /N_{0}$ to $%
\left( i_{1}+1,i_{2}\right) /N_{0}$. The contour integral on this segment
becomes%
\begin{eqnarray}
&&\int_{i_{1}/N_{0}}^{\left( i_{1}+1\right) /N_{0}}\frac{i}{2\pi }\mathrm{tr}%
\left( \bar{V}^{\dag }\left( k^{1},i_{2}/N_{0}\right) \partial _{\mu }\bar{V}%
\left( k^{1},i_{2}/N_{0}\right) \right) \,\mathrm{d}k^{1}  \notag \\
&=&\frac{i}{2\pi }\mathrm{tr}\ln \left( R^{\left( 1\right) }L^{\left(
1\right) \dag }\right)   \notag \\
&=&\,\frac{i}{2\pi }\ln \det \left( R^{\left( 1\right) }L^{\left( 1\right)
\dag }\right)   \notag \\
&=&\,\frac{i}{2\pi }\ln \frac{\det \left( R^{\left( 1\right) }\cos \Theta
^{\left( 1\right) }L^{\left( 1\right) \dag }\right) }{\det \cos \Theta }
\notag \\
&=&\frac{i}{2\pi }\ln \frac{\det \left( V^{\dag }\left( \left(
i_{1}+1\right) /N_{0},i_{2}/N_{0}\right) V\left( \left( i_{1}+1\right)
/N_{0},i_{2}/N_{0}\right) \right) }{\left\vert \det \left( V^{\dag }\left(
\left( i_{1}+1\right) /N_{0},i_{2}/N_{0}\right) V\left( \left(
i_{1}+1\right) /N_{0},i_{2}/N_{0}\right) \right) \right\vert }
\end{eqnarray}

which coincides with Eq.~(16) in \cite{Fukui2005}. In this way, our approach
provides an alternative derivation of the algorithm in \cite{Fukui2005} that
can be generalized to other topological invariants.

\section{Proof of the uniqueness of $\tilde{\rho}( \mathbf{k}) $}

Given $\boldsymbol{\kappa }_{0},\boldsymbol{\kappa }_{1}\in K_{0}$ such that
$C_{\boldsymbol{\kappa }_{0}}\cap C_{\boldsymbol{\kappa }_{1}}$ is
non-empty, we are going to show $V_{\boldsymbol{\kappa }_{0}}^{\left(
0\right) }\left( \mathbf{k}\right) \simeq V_{\boldsymbol{\kappa }%
_{1}}^{\left( 0\right) }\left( \mathbf{k}\right) $ for any $\mathbf{k}\in C_{%
\boldsymbol{\kappa }_{0}}\cap C_{\boldsymbol{\kappa }_{1}}$ by deduction.

\begin{figure}[tbh]
\centering\includegraphics[width=0.45\textwidth]{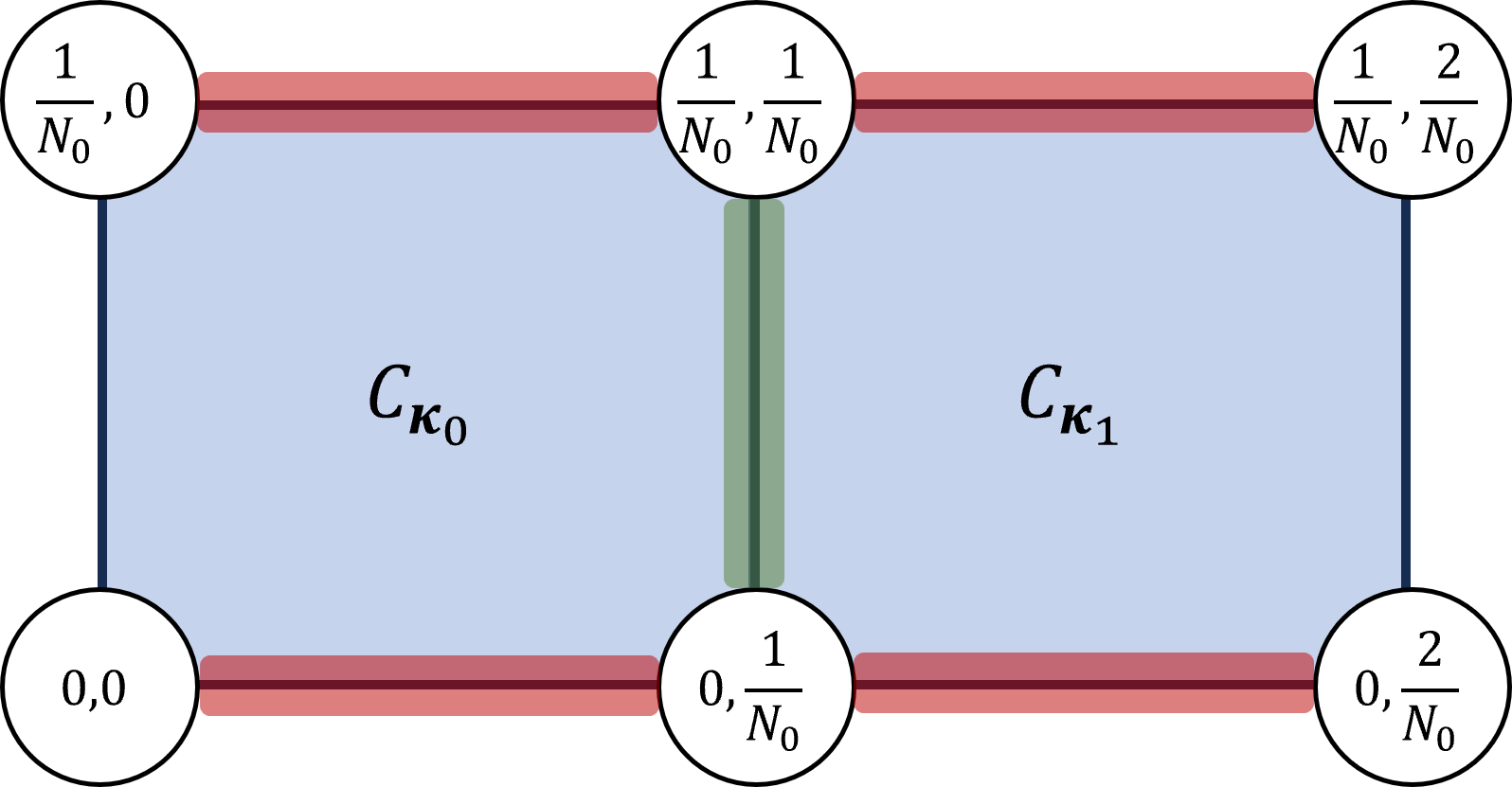}
\caption{ (Color online) A 2D grid to be interpolated}
\label{FigSM}
\end{figure}

First, we illustrate the idea of the proof for the case $D=2$. In Fig.~\ref%
{FigSM}, we suppose $C_{\boldsymbol{\kappa }_{0}}$ and $C_{\boldsymbol{%
\kappa }_{1}}$ are adjacent squares that share an edge. $V_{\boldsymbol{%
\kappa }_{0}}^{\left( 2\right) }\left( \mathbf{k}\right) \ $and $V_{%
\boldsymbol{\kappa }_{1}}^{\left( 2\right) }\left( \mathbf{k}\right) $ are
known for $\mathbf{k}$ at the vertices and their values are simply $V\left(
\mathbf{k}\right) $. $V_{\boldsymbol{\kappa }_{0}}^{\left( 1\right) }\left(
\mathbf{k}\right) \ $and $V_{\boldsymbol{\kappa }_{1}}^{\left( 1\right)
}\left( \mathbf{k}\right) $ are obtained as interpolants on the red edges of
the squares. In particular, $V_{\boldsymbol{\kappa }_{0}}^{\left( 1\right)
}\left( 0,1/N_{0}\right) \simeq V_{\boldsymbol{\kappa }_{0}}^{\left(
2\right) }\left( 0,1/N_{0}\right) =V_{\boldsymbol{\kappa }_{1}}^{\left(
2\right) }\left( 0,1/N_{0}\right) =V_{\boldsymbol{\kappa }_{1}}^{\left(
1\right) }\left( 0,1/N_{0}\right) $, so $V_{\boldsymbol{\kappa }%
_{0}}^{\left( 1\right) }\left( 0,1/N_{0}\right) \simeq V_{\boldsymbol{\kappa
}_{1}}^{\left( 1\right) }\left( 0,1/N_{0}\right) $. For the same reason, $V_{%
\boldsymbol{\kappa }_{0}}^{\left( 1\right) }\left( 1/N_{0},1/N_{0}\right)
\simeq V_{\boldsymbol{\kappa }_{1}}^{\left( 1\right) }\left(
1/N_{0},1/N_{0}\right) $. In the next step of interpolation, on the green
edge, $V_{\boldsymbol{\kappa }_{i}}^{\left( 0\right) }\left( \mathbf{k}%
\right) $ is determined by $V_{\boldsymbol{\kappa }_{i}}^{\left( 1\right)
}\left( 0,1/N_{0}\right) $ and $V_{\boldsymbol{\kappa }_{i}}^{\left(
1\right) }\left( 1/N_{0},1/N_{0}\right) $ for $i=0$ or $1$. Then $V_{%
\boldsymbol{\kappa }_{0}}^{\left( 0\right) }\left( \mathbf{k}\right) \simeq
V_{\boldsymbol{\kappa }_{1}}^{\left( 0\right) }\left( \mathbf{k}\right) $ on
the green edge follows from the uniqueness of the geodesic, because they
connect the same set of two hyperplanes.

Now consider the general case. Define $\boldsymbol{\xi} \equiv \boldsymbol{%
\kappa }_{1}-\boldsymbol{\kappa }_{0}$. Because $C_{\boldsymbol{\kappa }%
_{0}}\cap C_{\boldsymbol{\kappa }_{1}}$ is non-empty, $\xi ^{i}$ can only be
$0$ or $\pm N_{0}^{-1}$. We denote the intersection of the domains of $V_{%
\boldsymbol{\kappa }_{0}}^{\left( i\right) }\left( \mathbf{k}\right) $ and $%
V_{\boldsymbol{\kappa }_{1}}^{\left( i\right) }\left( \mathbf{k}\right) $ as
$I^{\left( i\right) }$. First, by definition we have $V_{\boldsymbol{\kappa }%
_{0}}^{\left( D\right) }\left( \mathbf{k}\right) =V_{\boldsymbol{\kappa }%
_{1}}^{\left( D\right) }\left( \mathbf{k}\right) =V\left( \mathbf{k}\right) $
for $\mathbf{k}\in I^{\left( D\right) }$. Then, suppose $V_{\boldsymbol{%
\kappa }_{0}}^{\left( i\right) }\left( \mathbf{k}\right) \simeq V_{%
\boldsymbol{\kappa }_{1}}^{\left( i\right) }\left( \mathbf{k}\right) $ for $%
\mathbf{k}\in I^{\left( i\right) }$. If $\xi ^{i}=N_{0}^{-1}$ (or $%
-N_{0}^{-1}$), we have $k^{i}=\kappa ^{1}$ (or $\kappa ^{0}$) for $\mathbf{k}%
\in I^{\left( i-1\right) }$, in other words, $\mathbf{k}=P_{\boldsymbol{%
\kappa }_{1},0}^{i}\left( \mathbf{k}\right) =P_{\boldsymbol{\kappa }%
_{0},1}^{i}\left( \mathbf{k}\right) $ (or $\mathbf{k}=P_{\boldsymbol{\kappa }%
_{0},0}^{i}\left( \mathbf{k}\right) =P_{\boldsymbol{\kappa }%
_{1},1}^{i}\left( \mathbf{k}\right) $). As a result, $V_{\boldsymbol{\kappa }%
_{0}}^{\left( i-1\right) }\left( \mathbf{k}\right) \simeq V_{\boldsymbol{%
\kappa }_{0}}^{\left( i\right) }\left( \mathbf{k}\right) $ and $V_{%
\boldsymbol{\kappa }_{1}}^{\left( i-1\right) }\left( \mathbf{k}\right)
\simeq V_{\boldsymbol{\kappa }_{1}}^{\left( i\right) }\left( \mathbf{k}%
\right) $, therefore $V_{\boldsymbol{\kappa }_{0}}^{\left( i-1\right)
}\left( \mathbf{k}\right) \simeq V_{\boldsymbol{\kappa }_{1}}^{\left(
i-1\right) }\left( \mathbf{k}\right) $. If $\xi ^{i}=0$, we have $P_{%
\boldsymbol{\kappa }_{1},0}^{i}\left( \mathbf{k}\right) =P_{\boldsymbol{%
\kappa }_{0},0}^{i}\left( \mathbf{k}\right) $ and $P_{\boldsymbol{\kappa }%
_{1},1}^{i}\left( \mathbf{k}\right) =P_{\boldsymbol{\kappa }%
_{0},1}^{i}\left( \mathbf{k}\right) $ for $\mathbf{k}\in I^{\left(
i-1\right) }$, so $V_{\boldsymbol{\kappa }_{1}}^{\left( i\right) }\left( P_{%
\boldsymbol{\kappa }_{1},0}^{i}\left( \mathbf{k}\right) \right) \simeq V_{%
\boldsymbol{\kappa }_{0}}^{\left( i\right) }\left( P_{\boldsymbol{\kappa }%
_{0},0}^{i}\left( \mathbf{k}\right) \right) $ and $V_{\boldsymbol{\kappa }%
_{1}}^{\left( i\right) }\left( P_{\boldsymbol{\kappa }_{1},1}^{i}\left(
\mathbf{k}\right) \right) \simeq V_{\boldsymbol{\kappa },0}^{\left( i\right)
}\left( P_{\boldsymbol{\kappa }_{0},1}^{i}\left( \mathbf{k}\right) \right) $%
. Then from Eq.~(2) we know $V_{\boldsymbol{\kappa }_{0}}^{\left(
i-1\right) }\left( \mathbf{k}\right) \simeq V_{\boldsymbol{\kappa }%
_{1}}^{\left( i-1\right) }\left( \mathbf{k}\right) $. So we have proved that
$V_{\boldsymbol{\kappa }_{0}}^{\left( 0\right) }\left( \mathbf{k}\right)
\simeq V_{\boldsymbol{\kappa }_{1}}^{\left( 0\right) }\left( \mathbf{k}%
\right) $ for $\mathbf{k}\in I^{\left( 0\right) }$ and the piecewise
function $\tilde{V}\left( \mathbf{k}\right) $ defines a unique interpolant
hyperplane $\tilde{\rho}\left( \mathbf{k}\right) \equiv \tilde{V}\left(
\mathbf{k}\right) \tilde{V}^{\dag }\left( \mathbf{k}\right) $ for $\mathbf{k}%
\in K$.

\section{Python code for the benchmark}

\begin{lstlisting}
# -*- coding: utf-8 -*-
"""
Benchmark the two algorithms Itp+CS3 and 4DItg that
calculate the 2nd Chern numbers of the
Dirac lattice model.

@author: Youjiang Xu
"""
from multiprocessing import Pool
from time import perf_counter

import numpy as np
import scipy.linalg as sla

PI2 = 2 * np.pi
PI22 = PI2 * PI2



###############################################################################
#Pauli matrices and Dirac matrices
sigma = [
    np.eye(2, dtype=complex),
    np.array([[0., 1.],[1., 0.]], dtype=complex),
    np.array([[0., -1j],[1j, 0.]], dtype=complex),
    np.array([[1., 0.],[0., -1.]], dtype=complex)
]
Gamma = [
    np.kron(sigma[3], sigma[0]),
    np.kron(sigma[1], sigma[1]),
    np.kron(sigma[1], sigma[2]),
    np.kron(sigma[1], sigma[3]),
    np.kron(sigma[2], sigma[0])
]



###############################################################################
#Some simple routines that will be used in the later part of the code
def uni(U):
    L, _, RH = sla.svd(U, full_matrices=False)
    return L @ RH

def mdot(V0, V1):
    return V0.T.conj() @ V1

def mdotR(V0, V1):
    return V0 @ V1.T.conj()

def logm_norm(A):
    w, v = sla.eig(A)
    return (v* (np.log(w)[None,:])) @ v.T.conj()

def logm_norm_batch(A):
    for ii in np.ndindex(A.shape[:-2]):
        A[ii] = logm_norm(A[ii])

def add_int2tuple(t : tuple, i : int, axis : int = 0):
    return t[:axis] + (t[axis]+i,) + t[axis+1:]

def measureFuncCall(func, *args):
    """
    Measure the time cost of the function call func(args)
    """
    start = perf_counter()
    ret = func(*args)
    end = perf_counter()
    print(f'{func.__name__} took {end - start:.4f} seconds')
    return ret



###############################################################################
#The function that gives the 2 lower energy states of the lattice Dirac model
def DiracLowerBands(m : float, c :float, k : np.ndarray):
    ds = np.empty((5,), dtype=float)
    ds[0] = m + c * np.sum(np.cos(PI2 * k))
    ds[1:] = np.sin(PI2 * k)
    H = sum(ds[i] * Gamma[i] for i in range(5))
    _, vH = sla.eigh(H)
    return vH[:,:2]



###############################################################################
#The functions in this part implements the new algorithm Itp+CS
def interpV1(V: np.ndarray, len_grid: int = 8, axis: int = -3):
    """
    Interpolate `V` along a single axis specified by `axis`.

    Parameters
    ----------
    V ((N1, ..., ND, dV, nV) complex ndarray) :
        `V[i1,...,iD]` represents `nV` orthonormal vectors of dimension `dV`
    len_grid (int) :
        V is interpolated at the points specified by
        `grid = np.linspace(0., 1., len_grid, endpoint=False)`.
        `V[i1...,i_axis:i_axis+2,...iD]` determines the interpolant
        `IV[i1...,i_axis*len_grid:(i_axis+1)*len_grid,...iD]`
    axis (int) :
        Indicates along which axis `V` is interpolated

    Returns
    -------
    IV ((N1, ..., Naxis * lg + 1, ..., ND, DH, numV) complex ndarray) :
        Interpolant of `V`

    """
    grid = np.linspace(0., 1., len_grid, endpoint=False)

    IV = np.empty(
        V.shape[:axis] + ((V.shape[axis] - 1)*len_grid + 1,) + V.shape[axis+1:],
        dtype=complex
    )

    for ii in np.ndindex(V.shape[:axis]):
        for kk in np.ndindex(V.shape[axis+1:-2]):
            IV[ii][0][kk] = V[ii][0][kk]
            for j in range(V.shape[axis]-1):
                V0 = IV[ii][j*len_grid][kk]
                V1 = V[ii][j+1][kk]
                L, S, RH = sla.svd(mdot(V1, V0))
                U1 = V1 @ L
                if (len_grid > 1):
                    Theta = np.arccos(np.clip(S, None, 1.))
                    csc = 1. / np.sin(Theta)
                    U0 = mdotR(V0, RH)
                    IV[ii + (slice(j*len_grid+1,(j+1)*len_grid),) + kk] = \
                    np.tensordot(
                        (U0[None,:,:]*(csc[None,:]
                        * np.sin(np.outer((1.-grid[1:]),Theta)))[:,None,:]
                        +U1[None,:,:]*(csc[None,:]
                        * np.sin(np.outer((   grid[1:]),Theta)))[:,None,:]),
                        RH,
                        1
                    )
                IV[ii][(j+1)*len_grid][kk] = U1 @ RH
    return IV


def interpVD(V: np.ndarray, len_grid_arr: np.ndarray):
    """
    Do the 1D interpolation `interpV1` along each axis of `V`

    Parameters
    ----------
    V ((N1, ..., ND, dV, nV) complex ndarray) :
        `V[i1,...,iD]` represents `nV` orthonormal vectors of dimension `dV`
    len_grid_arr (int ndarray) :
        `V` is interpolated along ith direction with the number of grid points
        given by `len_grid_arr[i]`.
        See 'interpV1'.

    Returns
    -------
    IV (..., Ni * len_grid_arr[i] + 1, ..., dV, nV) complex ndarray :
        Interpolant of V.
    """
    IV = V
    D = V.ndim - 2
    for axis in reversed(range(D)):
        IV = interpV1(IV, len_grid_arr[axis], axis)
    return IV


def calcA(V : np.ndarray, D : int = 3):
    """
    Calculate the Berry connection A
    """
    A = np.zeros((D,) + V.shape[:D] + (V.shape[-1],)*2, dtype=complex)
    for axis in range(D):
        shape_i = add_int2tuple(V.shape[:D], -1, axis)
        for jj in np.ndindex(shape_i):
            kk = add_int2tuple(jj, 1, axis)
            A[axis][jj] = logm_norm(uni(mdot(V[jj], V[kk])))
    return A


def calcCS3(A):
    CS3_shape = np.array(A.shape[1:4]) - np.ones(3, dtype=int)
    CS3 = np.empty(CS3_shape, dtype=float)
    for ii in np.ndindex(CS3.shape):
        ii0 = add_int2tuple(ii, 1, 0)
        ii1 = add_int2tuple(ii, 1, 1)
        ii2 = add_int2tuple(ii, 1, 2)
        ii01 = add_int2tuple(ii0, 1, 1)
        ii02 = add_int2tuple(ii0, 1, 2)
        ii12 = add_int2tuple(ii1, 1, 2)
        A0 = A[0][ii] + A[0][ii1] + A[0][ii2]+ A[0][ii12]
        A1 = A[1][ii] + A[1][ii0] + A[1][ii2]+ A[1][ii02]
        A2 = A[2][ii] + A[2][ii0] + A[2][ii1]+ A[2][ii01]
        p1A0 = A[0][ii1] - A[0][ii] + A[0][ii12] - A[0][ii2]
        p2A0 = A[0][ii2] - A[0][ii] + A[0][ii12] - A[0][ii1]
        p0A1 = A[1][ii0] - A[1][ii] + A[1][ii02] - A[1][ii2]
        p2A1 = A[1][ii2] - A[1][ii] + A[1][ii02] - A[1][ii0]
        p0A2 = A[2][ii0] - A[2][ii] + A[2][ii01] - A[2][ii1]
        p1A2 = A[2][ii1] - A[2][ii] + A[2][ii01] - A[2][ii0]
        CS3[ii] = np.trace(
                A0@(p1A2 - p2A1) + A1@(p2A0 - p0A2) + A2@(p0A1 - p1A0)
                ).real \
                + np.trace(A0 @ A1 @ A2).real / 2.
    return CS3 / (PI22 * 16.)

def calcCS3_A0is0(A):
    CS3_shape = np.array(A.shape[1:4]) - np.ones(3, dtype=int)
    CS3 = np.empty(CS3_shape, dtype=float)
    for ii in np.ndindex(CS3.shape):
        ii0 = add_int2tuple(ii, 1, 0)
        ii1 = add_int2tuple(ii, 1, 1)
        ii2 = add_int2tuple(ii, 1, 2)
        ii01 = add_int2tuple(ii0, 1, 1)
        ii02 = add_int2tuple(ii0, 1, 2)
        A1 = A[1][ii] + A[1][ii0] + A[1][ii2]+ A[1][ii02]
        A2 = A[2][ii] + A[2][ii0] + A[2][ii1]+ A[2][ii01]
        p0A1 = A[1][ii0] - A[1][ii] + A[1][ii02] - A[1][ii2]
        p0A2 = A[2][ii0] - A[2][ii] + A[2][ii01] - A[2][ii1]
        CS3[ii] = np.trace(A2 @ p0A1- A1 @ p0A2).real
    return CS3 / (PI22 * 16.)

def intCS3_block(
        Vblock : np.ndarray,
        len_grid : int = 8,
        IV0 : np.ndarray = None, IV1 : np.ndarray = None,
        flag_greedy : bool = True):
    """Integrate the Chern-Simons form in the boundary of a 4D cube
    Args:
        Vblock ((2,2,2,2,dV,nV) complex ndarray):
            `V[i1,...,iD]` represents `nV` orthonormal vectors of dimension `dV`

        len_grid (int, optional):
            The length of the grid fed to `interp1D`. Defaults to 8.

        IV0: 3D interpolants of V[0]

        IV1: 3D interpolants of V[1]

        flag_greedy (bool) : if `False`, simply integrate the Chern-Simons form;
            if `True`, use the faster method that
            for some parts of the boundaries
            calculates the sum of the two integrals contributed by
            two adjacent 4D cubes, therefore the result is meaningful
            only when the results from all 4D cubes are summed up

    Returns:
        float : The integrated Chern-Simons form.
    """
    dim = 3

    grid3D = np.array((len_grid,)*3)
    if IV0 is None:
        IV0 = interpVD(Vblock[0], grid3D)
    if IV1 is None:
        IV1 = interpVD(Vblock[1], grid3D)
    IV2 = np.empty_like(IV0)
    for ii in np.ndindex(IV0.shape[:-2]):
        IV2[ii] = IV1[ii] @ uni(mdot(IV1[ii], IV0[ii]))

    A0 = calcA(IV0)
    A2 = calcA(IV2)

    ret = np.sum(calcCS3(A2) - calcCS3_A0is0(A0))

    if flag_greedy:
        coef = 1./(PI22 * 8.)
        for i in range(dim):
            iv = np.take(IV0, -1, i)
            ivg = interpVD(Vblock[0][(slice(None),)*i + (-1,)], (len_grid,)*2)
            g = np.einsum('ijkl,ijkm->ijlm', iv.conj(), ivg)
            dgg = np.zeros((2,)+g.shape, dtype=complex)
            dgg[0, :-1]  = np.einsum('ijkl,ijml->ijkm', g[1:], g[:-1].conj())
            logm_norm_batch(dgg[0, :-1])
            dgg[1,:,:-1] = np.einsum('ijkl,ijml->ijkm', g[:,1:], g[:,:-1].conj())
            logm_norm_batch(dgg[1,:,:-1])

            diff_a = np.delete(A0[(slice(None),)*(i+1) + (-1,)], i, 0) \
                - np.delete(A2[(slice(None),)*(i+1) + (-1,)], i, 0)
            ret += coef * sum(
                (
                np.trace(
                    (dgg[0][ii] + dgg[0][add_int2tuple(ii,1,1)])
                    @(diff_a[1][ii] + diff_a[1][add_int2tuple(ii,1,0)])
                    -(dgg[1][ii] + dgg[1][add_int2tuple(ii,1,0)])
                    @(diff_a[0][ii] + diff_a[0][add_int2tuple(ii,1,1)])
                ).real

                for ii in np.ndindex((dgg.shape[1]-1, dgg.shape[2]-1))
                )
            )
            coef = -coef
    else:
        coef = 1.
        for i in range(dim):
            iv = interpV1(
                np.stack((np.take(IV0, (0,-1), i), np.take(IV1, (0,-1), i)), axis=0),
                len_grid, 0)
            ret += coef * np.sum(
                calcCS3_A0is0(calcA(np.take(iv, 0, i+1)))
              - calcCS3_A0is0(calcA(np.take(iv,-1, i+1))))
            coef = -coef

    return ret, IV0, IV1

def intCS3_stack(Vstack : np.ndarray, len_grid : int = 8, flag_greedy : bool = True):
    """Integrate the Chern-Simons form in a stack of 4D cubes

    Args:
        Vstack ((n_block,2,2,2,dV,nV) complex ndarray):
            `V[i1,...,iD]` represents `nV` orthonormal vectors of dimension `dV`

        len_grid (int, optional): The length of the grid fed to `interp1D`.
            Defaults to 8.

        flag_greedy (bool) : See `intCS3_block`

    Returns:
        float : The integrated Chern-Simons form.
    """
    n_block = Vstack.shape[0] - 1
    ret, IV0, IV1 = intCS3_block(Vstack[0:2], len_grid, flag_greedy=flag_greedy)
    for i in range(1, n_block-1):
        tmp, _, IV1 = \
        intCS3_block(Vstack[i:i+2], len_grid, IV1, flag_greedy=flag_greedy)
        ret += tmp
    if n_block > 1:
        if flag_greedy:
            tmp, _, _ = \
            intCS3_block(Vstack[-2:], len_grid, IV1, IV0, flag_greedy=flag_greedy)
        else:
            tmp, _, _ = \
            intCS3_block(Vstack[-2:], len_grid, IV1, flag_greedy=flag_greedy)
        ret += tmp
    return ret


def intCS3(V : np.ndarray, len_grid : int = 8, flag_greedy : bool = True):
    """Integrate the Chern-Simons form in the whole parameter space
    """
    shape_stack = tuple(np.array(V.shape[1:4], dtype=int) - np.ones(3, dtype=int))
    ret = np.empty(shape_stack, dtype=float)
    args = (
        (V[:, ii[0]:ii[0]+2, ii[1]:ii[1]+2, ii[2]:ii[2]+2], len_grid, flag_greedy)
        for ii in np.ndindex(shape_stack)
    )
    with Pool() as p:
        for ii, tmp in zip(np.ndindex(shape_stack), p.starmap(intCS3_stack, args)):
            ret[ii] = tmp
    return np.sum(ret)



###############################################################################
#The three functions in this part are used to calculate the 2nd Chern number
#by the algorithm proposed in
#M Mochol-Grzelak et al 2019 Quantum Sci. Technol. 4 014009
def intF2_A(A : np.ndarray):
    grid = tuple(np.array(A.shape[1:5], dtype=int) - np.ones(4, dtype=int))
    F = np.empty((6,) + A.shape[-2:], dtype=complex)
    components = [(0,1),(0,2),(0,3),(1,2),(1,3),(2,3)]

    ret = 0.
    for jj in np.ndindex(grid):
        kk = [add_int2tuple(jj, 1, i) for i in range(4)]
        for i, c in enumerate(components):
            F[i] = A[c[0]][jj] @ A[c[1]][jj]
            F[i] -= F[i].T.conj()
            F[i] += A[c[1]][kk[c[0]]] - A[c[1]][jj] \
                   -A[c[0]][kk[c[1]]] + A[c[0]][jj]
        ret += np.trace(F[0]@F[5] - F[1]@F[4] + F[2]@F[3])
    return ret / PI22

def calc_F2_U(Ublock : np.ndarray):
        idx_pair = [(0,1),(0,2),(0,3),(1,2),(1,3),(2,3)]
        F = np.empty((6,) + Ublock.shape[-2:], dtype=complex)
        t0 = (0,0,0,0)
        for i, ip in enumerate(idx_pair):
            F[i] = Ublock[ip[0]][t0] \
                @ Ublock[ip[1]][add_int2tuple(t0,1,ip[0])] \
                @ sla.inv(Ublock[ip[0]][add_int2tuple(t0,1,ip[1])]) \
                @ sla.inv(Ublock[ip[1]][t0])
            F[i] = sla.logm(F[i])
        return np.trace(F[0]@F[5] - F[1]@F[4] + F[2]@F[3]).real

def intF2(V : np.ndarray):
    U = np.zeros((4,) + V.shape[:4] + (V.shape[-1],)*2, dtype=complex)
    for i in range(4):
        U_non0_shape = V.shape[:i] + (V.shape[i]-1,) + V.shape[i+1:4]
        for jj in np.ndindex(U_non0_shape):
            U[i][jj] = mdot(V[jj], V[add_int2tuple(jj, 1, i)])

    grid = tuple(np.array(V.shape[:4], dtype=int) - np.ones(4, dtype=int))
    args = [
        U[:,jj[0]:jj[0]+2,jj[1]:jj[1]+2,jj[2]:jj[2]+2,jj[3]:jj[3]+2]
        for jj in np.ndindex(grid)
    ]
    with Pool() as p:
        return sum(p.map(calc_F2_U, args)) / PI22



###############################################################################
if __name__ == "__main__":
    def benchmark_4DInt(N, len_grid, m, c):
        N *= len_grid
        V = np.empty((N+1,)*4 + (4, 2), dtype=complex)
        args = ((m, c, np.array(ii, dtype=float)/N) for ii in np.ndindex((N,)*4))
        with Pool() as p:
            for ii, ret in zip(np.ndindex((N,)*4), p.starmap(DiracLowerBands, args)):
                V[ii] = ret
        for i in range(4):
            V[(slice(None),)*i + (-1,)] = V[(slice(None),)*i + (0,)]
        rst = intF2(V)
        return rst, rst - round(rst), abs(1 - round(rst)/rst)

    def benchmark_ItpCS(N, len_grid, m, c, flag_greedy : bool = True):
        V = np.empty((N+1,)*4 + (4, 2), dtype=complex)
        args = ((m, c, np.array(ii, dtype=float)/N) for ii in np.ndindex((N,)*4))
        with Pool() as p:
            for ii, ret in zip(np.ndindex((N,)*4), p.starmap(DiracLowerBands, args)):
                V[ii] = ret
        for i in range(4):
            V[(slice(None),)*i + (-1,)] = V[(slice(None),)*i + (0,)]
        rst = intCS3(V, len_grid, flag_greedy)
        return rst, rst - round(rst), abs(1 - round(rst)/rst)

    N = 4
    len_grid_arr = [4, 6, 8, 10]
    m_arr = [1.0, 0.1, 0.01, 3.0, 3.9, 3.99]
    c = 1.

    for m in m_arr:
        for len_grid in len_grid_arr:
            print(f"m = {m:}, N1 = {len_grid:}")
            print(measureFuncCall(benchmark_ItpCS, N, len_grid, m, c, True))
            print(measureFuncCall(benchmark_4DInt, N, len_grid, m, c))
\end{lstlisting}

\end{document}